\documentclass[showpacs,aps,prd,nofootinbib,floatfix,amsmath,amssymb]{revtex4}
\usepackage{graphicx}

\begin{document}


\title{$H^\pm W^\mp Z$ contribution to the static quantities of the $W$ boson in the context of Higgs-triplet theories}
\author{G. Tavares--Velasco}
\email[E-mail:]{gtv@fcfm.buap.mx}
\author{J. J. Toscano}
\email[E-mail:]{jtoscano@fcfm.buap.mx} \affiliation{Facultad de
Ciencias F\'\i sico Matem\' aticas, Benem\' erita Universidad
Aut\' onoma de Puebla, Apartado Postal 1152, Puebla, Pue.,
M\'exico}

\date{\today}

\begin{abstract}
We calculate the one--loop contribution from the $H^\pm W^\mp Z$
coupling to the static electromagnetic properties of the $W$
boson. Although this coupling is absent at the tree--level in all
Higgs--doublet models, it can be induced at this order in models
including Higgs--triplet representations. It is found that the
$H^\pm W^\mp Z$ contribution can be as important as those arising
from other couplings including Higgs bosons, such as the standard
model coupling $W^\pm W^\mp H$ or the two--Higgs--doublet model
couplings $H^\pm W^\mp \phi^0$ and $W^\pm W^\mp \phi^0$, with
$\phi^0=h,\,H$ and $A$.

\end{abstract}


 \maketitle

\section{Introduction}
\label{int} The trilinear gauge boson couplings $W^\pm W^\mp
\gamma$ and $W^\pm W^\mp Z$ are representative of the nonabelian
structure of the standard model (SM). It is thus interesting to
study any anomalous (one--loop) contribution to them as it is
important to test the quantization procedures used for these
nonabelian gauge systems. In this context, the CP--even static
electromagnetic properties of the $W$ boson, which are
parametrized by two form factors, $\Delta \kappa$ and $\Delta Q$,
have been the subject of considerable interest in the literature
as is expected that future particle colliders be sensitive to this
class of effects \cite{Diehl}. The SM contributions to $\Delta
\kappa$ and $\Delta Q$ were calculated in Ref. \cite{Bardeen} for
massless fermions, and the top quark contribution was presented in
\cite{Top}. The sensitivity of these quantities to new physics
effects has also been analyzed in some SM extensions, such as the
two--Higgs--doublet model (THDM) \cite{THDM}, supersymmetric
theories \cite{SUSY}, left--right symmetric models \cite{Roberto},
$SU(3)\times SU(3)\times U(1)$ models\cite{TT}, and models with
composite particles \cite{Rizzo} and an extra gauge boson
$Z^\prime$ \cite{Sharma}. $\Delta \kappa$ and $\Delta Q$ have also
been parametrized in a model independent manner by using the
effective Lagrangian technique \cite{EL}. In this work we present
the calculation of the contribution from the $H^\pm W^\mp Z$
coupling to the static quantities of the $W$ boson. Charged Higgs
bosons appear in many extensions of the SM, such as the popular
THDM. In the search for a charged Higgs boson, the $H^\pm W^\mp Z$
coupling might play an important role, although it is expected to
be very suppressed in a Higgs--doublet model. In fact, although
the $H^\pm W^\mp Z$ coupling can have a renormalizable structure,
it can only be generated at the one-loop level in
multi--Higgs--doublet models \cite{HWZ}. Nevertheless, it can be
induced at the tree--level in theories with Higgs triplets or
higher representations, though it could be severely constrained by
the $\rho$ parameter. This is true in some Higgs--triplet models
which do not respect the $SU(2)$ custodial symmetry \cite{HHG,GM}.
It is possible however to construct a model including Higgs
triplets that does respect such a custodial symmetry
\cite{G,CG,GVW}, thereby relaxing the constraints from the $\rho$
parameter. The phenomenology of the $H^\pm W^\mp Z$ coupling has
been investigated in the context of the CERN $e^+e^-$ LEP--II
collider, the next linear collider \cite{LC}, and hadronic
colliders \cite{HC}. Such studies focus mainly on discriminating a
charged Higgs bosons arising from a model with Higgs triplets from
that induced  by a Higgs--doublet model. The main goal of this
work is to study the impact of the $H^\mp W^\mp Z$ vertex on the
static electromagnetic properties of the $W$ boson. We will show
that the respective contributions to $\Delta \kappa$ and $\Delta
Q$ may be of the same order of magnitude as those arising from
other renormalizable theories which include neutral o charged
Higgs bosons, such as the SM or the THDM.

The rest of the paper is organized as follows. In Sec. \ref{am} we
present the calculation of the $\Delta \kappa$ and $\Delta Q$ form
factors. Sec. \ref{dis} is devoted to discuss the numerical
results. Finally, the conclusions are presented in Sec. \ref{c}.

\section{$H^\pm W^\mp Z$ contribution to the on--shell $W^\pm W^\mp \gamma$ vertex}
\label{am}Using the momenta depicted in Fig. 1, the most general
CP--even Lorentz structure for the on--shell $W^\pm W^\mp \gamma$ vertex can
be written as

\begin{equation}
\label{eq1} \Gamma_{\alpha \beta \mu}=ie\left(A\left(2p_\mu
g_{\alpha \beta}+4(q_\beta g_{\alpha \mu}-q_\alpha g_{\beta
\mu})\right)+2\Delta \kappa (q_\beta g_{\alpha \mu}-q_\alpha
g_{\beta \mu})+\frac{4\Delta Q}{m^2_W}p_\mu q_\alpha
q_\beta\right).
\end{equation}

\begin{figure}
\centering
\includegraphics[width=3in]{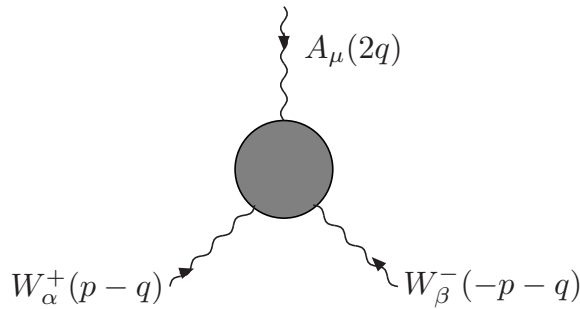}
\caption{Trilinear $W^\pm W^\mp \gamma$ coupling. The loop denotes
contributions from charged particles.}
\end{figure}

In the SM, both $\Delta \kappa$ and $\Delta Q$ vanish at the
tree--level, whereas the one--loop corrections are of the order of
$\alpha/\pi$. These parameters define the magnetic dipole moment
$\mu_W$ and the electric quadrupole moment $Q_W$ of the $W$ boson:
\begin{eqnarray}
\mu_W&=&\frac{e}{2m_W}(2+\Delta \kappa), \\
Q_W&=&-\frac{e}{m^2_W}(1+\Delta \kappa+\Delta Q).
\end{eqnarray}

It is worth discussing the origin of the $H^\pm W^\mp Z$ vertex.
It has the following renormalizable structure which is dictated by
Lorentz covariance
\begin{equation}
i\,s_H\,g\,m_Z\,g_{\mu \nu},
\end{equation}
where $s_H$ is a model--dependent quantity which may be very
suppressed in models containing only Higgs doublets \cite{HWZ}. As
pointed out before, in models with a scalar sector including
doublets and triplets, $s_H$ may be considerably enhanced since
the $H^\pm W^\mp Z$ vertex is induced at the tree--level. However,
not any scalar sector including triplets or higher representations
is viable due to the fact that large deviations from the
tree--level relation $\rho=1$ may arise. One possibility is to
invoke a tree--level custodial $SU(2)$ symmetry respected by the
Higgs sector, which guarantees that $\rho=1$ at the tree--level.
In this case, the existence of a tree--level--induced $H^\pm W^\mp
Z$ vertex with strength $s_H$ of the order of unity is possible.
Several models of this class have been proposed in the literature,
but we will focus on that introduced by Georgi {\it et. al.}
\cite{G}, and later considered in \cite{CG} and \cite{GVW}. The
Higgs sector of such a model consists of a complex doublet with
hypercharge $Y=1$, a real triplet with $Y=0$, and a complex
triplet with $Y=2$ \cite{G,GVW}. In this model $s_H$ is the sine
of a doublet--triplet mixing angle and is given by
\begin{equation}
s_H=\sqrt{\frac{8w^2}{v^2+8w^2}},
\end{equation}
where $v$ is the vacuum expectation value (VEV) of the Higgs
doublet and $w$ is that of both Higgs triplets \cite{GVW}. As is
evident from the above expression, there are two extreme scenarios
which have direct implication on the $H^\pm W^\mp Z$ vertex. One
scenario corresponds to the case in which the spontaneous symmetry
breaking (SSB) of the electroweak sector is entirely determined by
the Higgs doublet, {\it i.e.}, $v\gg w$, implying that $s_H\sim
0$, which means that the $H^\pm W^\mp Z$ vertex is strongly
suppressed. The more promising scenario for having $s_H$ of order
$O(1)$ corresponds to the case where $w\gg v$, which means that
the SSB of the theory is dictated by the Higgs triplet. This
scenario is very appealing and so it will be considered below.

We turn now to present the calculation of the contribution from
the $H^\pm W^\mp Z$ vertex to $\Delta \kappa$ and $\Delta Q$. This
contribution is given by the Feynman diagram shown in Fig. 2. In
the unitary gauge, the respective amplitude can be written as
\begin{equation}
\label{gamma}
\Gamma_{\alpha \beta
\mu}=-2\,e\,g^2\,m^2_Z\,s^2_H\int\frac{d^Dk}{(2\pi)^D}\frac{(k+p)_\mu
(g_{\alpha \beta}-\frac{1}{m^2_Z}k_\alpha
k_\beta)}{[k^2-m^2_Z][(k+p-q)^2-m^2_{H^\pm}][(k+p+q)^2-m^2_{H^\pm}]}.
\end{equation}
Once the Feynman parameters technique is applied, we obtain the
following expressions for the electromagnetic form factors of the
$W$ boson

\begin{eqnarray}
\label{DKparam}
\Delta \kappa&=&6\,a\,s^2_H\int^1_0 dx\int^{1-x}_0  dy\left((1-2\,x-y)\log {\cal R}-\frac{2\,m_Z^2(1-x-y)}{\cal R}\right),\\
\label{DQparam}
 \Delta Q&=&12\,a\,s^2_H\int^1_0 dx
\int^{1-x}_0  dy\frac{\,m_W^2(1-x-y)x\, y}{\cal R}.
\end{eqnarray}
with $a=g^2/(96\pi^2)$ and ${\cal
R}=m_Z^2-\left(m_W^2+m_Z^2-m_{H^\pm}^2\right)(x+y)+m_W^2(x+y)^2$.
After some calculation one ends up with the explicit solution

\begin{eqnarray}
\label{DK} \Delta
\kappa&=&a\,s^2_H\left(A_1(x_h,x_z)+A_2(x_h,x_z)\log\left(\frac{x_z}{x_h}\right)+A_3(x_h,x_z)\frac{f(x_h,x_z)}{\delta(x_h,x_z)}\right),\\
\label{DQ} \Delta
Q&=&a\,s^2_H\left(B_1(x_h,x_z)+B_2(x_h,x_z)\log\left(\frac{x_z}{x_h}\right)+B_3(x_h,x_z)\frac{f(x_h,x_z)}{\delta(x_h,x_z)}\right),
\end{eqnarray}
where  the $A_i$ and $B_i$ functions are given by
\begin{eqnarray}
A_1(x,y)&=&\frac{1}{2}\left(1+6(x^2-y^2)^2+3(7y^2-3x^2)\right),
\\
A_2(x,y)&=&3\left((x^2-y^2)^3-2(x^2-y^2)^2+y^2(x^2-y^2)+x^2\right),
\\
A_3(x,y)&=&\frac{3}{2}\left((x^2-y^2)^4-3(x^2-y^2)^3-
y^2(x^2-y^2)^2+3(x^4-y^4)-x^2(1+4y^2)\right),
\\
B_1(x,y)&=&-\frac{2}{3}-2(x^2-y^2)^2+3x^2-y^2,
\\
B_2(x,y)&=&-2\left((x^2-y^2)^3-2x^2(x^2-y^2)+x^2\right),\\
B_3(x,y)&=&-(x^2-y^2)^4+3x^4(x^2-1)+y^4(x^2+y^2)-x^2y^2(5x^2-1)+x^2.
\end{eqnarray}
In addition, we introduced the definitions $x_h=m_{H^\pm}/m_W$ and
$x_z=m_Z/m_W$, along with
\begin{eqnarray}
\delta(x,y)&=&\sqrt{1-2(x^2+y^2)+(x^2-y^2)^2}, \\
f(x,y)&=&\log\left(\frac{1-(x^2+y^2)-\delta(x,y)}{1-(x^2+y^2)+\delta(x,y)}\right).
\end{eqnarray}

\begin{figure}
\centering
\includegraphics[width=2.5in]{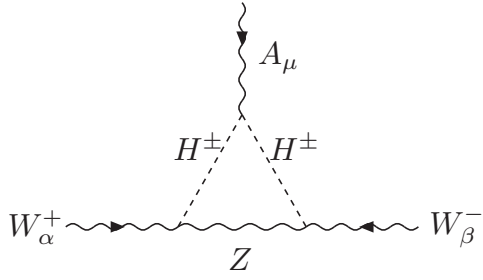}
\caption{$H^\pm W^\mp Z$ contribution to the on-shell $W^\pm W^\mp \gamma$
vertex.}
\end{figure}

In order to cross--check the above results, $\Delta \kappa$ and
$\Delta Q$ were calculated independently by a general method for
reducing tensor form factors. This method, which is an extension
of the Passarino--Veltman scheme \cite{Veltman2}, is described
detailed in Ref. \cite{Stuart}. In such a scheme one assumes that
$q^2\ne 0$ and applies the usual Passarino--Veltman reduction. The
reason why one cannot put $q^2=0$ in Eq. (\ref{gamma}) prior to
applying the tensor reduction is that it would require the
inversion of a kinematic matrix whose determinant is
$\|{\mathbf{\cal D}}\|=4\,q^2\left(m^2_W-q^2\right)$, which
evidently vanishes for $q^2=0$. Once the form factors for
arbitrary $q^2$ are obtained, the limit $q^2\to 0$ is taken in
order to yield the static quantities of the $W$ boson in terms of
two--point Passarino-Veltman scalar functions $B_0$. Although the
limiting procedure $q^2\to 0$ involves some additional
complications since the application of L'H\^{o}pital rule is
required, one important advantage of this method is that it can be
computer programmed, thereby eliminating the possibility of any
mistake. After this scheme is applied, one is left with the
following results:

\begin{eqnarray}
\frac{\Delta \kappa}{a\,s^2_H}&=&\frac{1}{2} - \frac{3}{2}\,x_z^2 \,\left( 9 +
2\,x_z^2 \right) + \frac{3}{2}\,x_h^2\left(1  + 4\,x_h^2
-2\,x_h^2\right) + \frac{24\,x_z^2}{\delta^2(x_h,x_z)} \,\left( 1
- x_z^2 -x_h^2 \right)+ \frac{3 \,x_h^2}{\delta^2(x_h,x_z)}{\left(
1 + x_z^2 - x_h^2 \right) }^3\, \Delta B_1\nonumber\\ &-&3\,x_z^2 \,
\left(x_z^2 - x_h^2  - \frac{4\,x_z^2}{\delta^2(x_h,x_z)} \,\left(
1 - x_z^2 + x_h^2 \right) \right)\,\Delta B_2,
\end{eqnarray}

\begin{eqnarray}
\frac{\Delta Q}{a\,s^2_H}&=& -\frac{2}{3} +x_z^2 \left(3  + 2\,x_z^2\right) -
x_h^2\left( 1 + 4\,x_z^2 -2\,x_h^2\right)  -
\frac{4\,x_z^2}{\delta^2(x_h,x_z)} \,\left( 1 - x_z^2 - x_h^2
\right) \nonumber\\&-& \frac{2\,x_h^2}{\delta^2(x_h,x_z)} \,
\left(x_z^6 +{\left( 1 - x_h^2 \right) }^3 -3\,\left( 1 -
x_h^2 +x_z^2 \right) \,x_z^2 \,x_h^2 \right) \,{\Delta B_1}\nonumber\\
&+& \frac{2\,x_z^2}{\delta^2(x_h,x_z)} \, \left( \left( x_h^2 -
x_z^2 \right) \, \left( \left(1 - x_z^2 +x_h^2\right) \,x_z^2 +
x_h^2\left(2+x_z^2 - x_h^2 \right) \right)
-x_h^2\right)\,{\Delta B_2}.
\end{eqnarray}

\noindent with $\Delta B_1=B_0(m_W^2, m_{H^\pm}^2, m_Z^2)-B_0(0,
m_{H^\pm}^2, m_{H^\pm}^2)$ and $\Delta B_2=B_0(m_W^2,
m_{H^\pm}^2,m_Z^2)-B_0(0, m_Z^2, m_Z^2)$. These scalar integrals
can be solved analytically or numerically evaluated \cite{FF}. We
compared numerically the latter results with those obtained from
Eqs. (\ref{DK})--(\ref{DQ}) and observed a perfect agreement. In
this way we make sure that our results are correct.

\section{Discussion}
\label{dis} 
\begin{figure}
\centering
\includegraphics[width=3in]{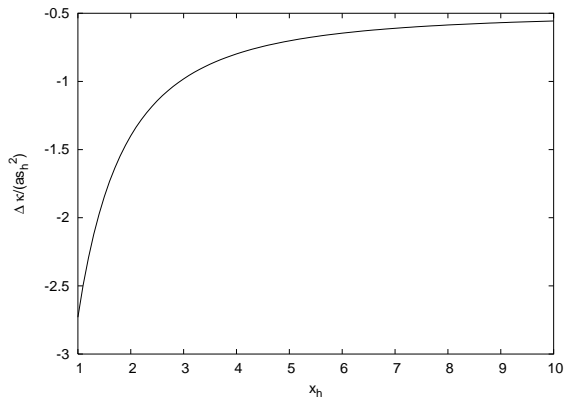}
\caption{\label{fig1} $\frac{\Delta \kappa}{a\,s^2_H}$ as a
function of $x_h=\frac{m_{H^\pm}}{m_W}$.}
 \end{figure}

\begin{figure}
\centering
\includegraphics[width=3in]{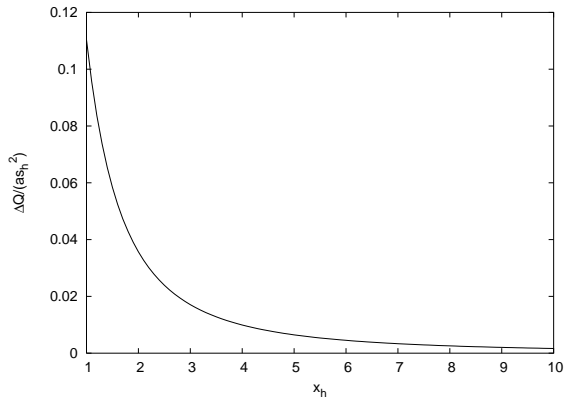}
\caption{\label{fig2} $\frac{\Delta Q}{a\,s^2_H}$ as a function of
$x_h=\frac{m_{H^\pm}}{m_W}$.}
 \end{figure}

The behavior of $\Delta \kappa/(a\,s^2_h)$ and $\Delta
Q(a\,s^2_H)$ as a function of $x_h$ is shown in Figs. \ref{fig1}
and \ref{fig2}, respectively. From these figures we can observe
that $\Delta \kappa$ is of the order of $a\,s^2_H$ in the range
$2\,m_W<m_{H^\pm}<10\,m_W$, whereas $\Delta Q$ is one order of
magnitude below. One can also observe that $\Delta Q$ decreases
more rapidly than $\Delta \kappa$ with increasing $m_{H^\pm}$. It
is convenient to compare our results with those arising from other
couplings involving Higgs bosons. For instance, the contribution
from the SM $W^\pm W^\mp H^0$ vertex was calculated in Ref. \cite{Bardeen},
in which case $\Delta \kappa\sim a$ and $\Delta Q\sim 10^{-2}a$
for values of $m_H$ in the same range considered for $m_{H^\pm}$
in Figs. \ref{fig1} and \ref{fig2}. Similar results were found in
Ref. \cite{THDM} for the contributions coming from the THDM
couplings $W^\pm W^\mp \phi^0$ and $H^\pm W^\mp\phi^0$, with
$\phi^0=H^0,\,h^0$, and $A^0$. We conclude thus that the
contribution of the $H^\pm W^\mp Z$ vertex to $\Delta \kappa$ and
$\Delta Q$ may be as important as those contributions arising from
other Higgs boson couplings provided that $s_H\sim 1$, {\it i.e},
when the SSB of the electroweak sector is dictated by the Higgs
triplets.

We now would like to focus on the decoupling nature of the Higgs
boson contributions to the $\Delta Q$ and $\Delta \kappa$ form
factors.  The sensitivity of $\Delta \kappa$ to heavy physics
effects as well as the decoupling nature of $\Delta Q$ was
analyzed in a more general context in Ref. \cite{Inami}. From
Figs. \ref{fig1} and \ref{fig2} we can see that $\Delta Q$
decouples for a large Higgs scalar mass, whereas $\Delta \kappa$
does not. These results do not contradict the decoupling theorem,
which establishes that those Lorentz structures arising from
renormalizable operators can be sensitive to nondecoupling
effects, whereas those structures coming from nonrenormalizable
operators are suppressed by inverse powers of the heavy mass,
thereby decoupling in the large mass limit. As far as  $\Delta Q$
and $\Delta \kappa$ are concerned, the former always decouples for
a large mass of a particle running in the loop since its Lorentz
structure is generated by a nonrenormalizable dimension--six
operator; on the contrary, $\Delta \kappa$ may be sensitive to
nondecoupling effects as the Lorentz structure associated with
this quantity is induced by a renormalizable dimension--four
operator. In this context, the decoupling properties of $\Delta Q$
and $\Delta \kappa$ have been discussed in the context of all of
the renormalizable theories considered up to now in the literature
\cite{Bardeen,TT}. For instance,  in the SM the heavy Higgs mass
limit yields $\Delta Q\to 0$ and $\Delta \kappa\to a$
\cite{Bardeen}. As far as the THDM is concerned, $\Delta \kappa\to
2\,a$ and $\Delta Q \to 0$ when $m_{\phi^0}$ becomes very large
and $m_{H^\pm}$ is kept fixed, whereas $\Delta \kappa\to -a$ and
$\Delta Q\to 0$ in the opposite scenario \cite{THDM}. In addition,
in the same model, both $\Delta \kappa$ and $\Delta Q$ vanish when
both $m_{\phi^0}$ and $m_{H^\pm}$ are very large. In our case, we
obtain that $\Delta \kappa\to - (a\, s_H^2)/2$ and $\Delta Q\to 0$
in the heavy charged scalar mass limit. These values follow
readily from Eq. (\ref{DKparam}).

It is interesting to note that our results can also be used to
obtain the contribution to the static quantities of the $W$ boson
from the $H^\mp W^\pm Z^{\prime}$ coupling, with $Z^\prime$ the
extra neutral gauge boson appearing in theories with an extra
$U(1)$. This contribution has yet been calculated in the context
of an extra $U(1)$ superstring inspired theory \cite{Sharma}.
According to these authors, their results were obtained in the
Landau gauge. We have compared numerically our results with those
presented Eqs. (12) and (13) of Ref. \cite{Sharma} and found no
agreement. As pointed out above, our results were cross--checked
by making a comparison between the results obtained via the
Feynman parameters technique and those obtained by the slightly
modified version of the Passarino--Veltman scheme described in
Refs. \cite{TT,Stuart}. These two methods are independent and
allows us to make sure that our results are correct.

\section{Conclusion}
\label{c}The novel feature of a Higgs--triplet representation is
the presence of a tree--level $H^\pm W^\mp Z$ vertex, whose
strength may be of the order of the unity provided that the SSB of
the electroweak sector is dictated by the VEV of the neutral
components of the Higgs triplet. This class of models can be
viable as long as a tree--level custodial $SU(2)$ symmetry is
respected by the Higgs potential. The phenomenology of the $H^\pm
W^\mp Z$ coupling  is thus be very appealing. Any direct or
indirect evidence of the this coupling would be a clear signal of
the existence of a scalar sector comprised by Higgs triplets. We
have studied the impact of this vertex on the static
electromagnetic properties of the $W$ gauge boson. We found that
the respective contributions to the form factors $\Delta \kappa$
and $\Delta Q$ are as important as those predicted by the SM
coupling $W^\pm W^\mp H^0$ or those arising from the THDM
couplings $W^\pm W^\mp\phi^0$ and $H^\pm W^\mp \phi^0$.

We would like to point out that our results can also be
used to evaluate the contributions from the $H^\pm W^\mp Z^\prime$
vertex to the magnetic moments of the $W$ boson, where $Z^\prime$
is the neutral boson which appears in theories with an extra
$U(1)$ gauge symmetry. We note that our results, (\ref{DK}) and
(\ref{DQ}), disagree with those presented in Ref. \cite{Sharma}
for the contribution from the $H^\pm W^\mp Z^\prime$ coupling in
an extra $U(1)$ superstring inspired theory. Finally, We emphasize that our
results were cross--checked by making a comparison between the
results obtained via the Feynman parameters technique and those
obtained by the slightly modified version of the
Passarino--Veltman scheme described in Ref. \cite{Stuart}.

\acknowledgments{Support from CONACYT and SNI (M\' exico) is
acknowledged. The work of G. T. V. is also supported by
SEP-PROMEP.}

\end{document}